\documentclass[12pt]{article}
\usepackage{amsmath}

\def\a{\alpha}
\def\b{\beta}

\def\d{\delta}

\def\g{\gamma}
\def\p{\psi}

\def\k{\kappa}
\def\la{\lambda}
\def\m{\mu}
\def\n{\nu}
\def\om{\omega}

\def\s{\sigma}

\def\wH{\widetilde{H}}
\def\wD{\widetilde{D}}
\def\wK{\widetilde{K}}

\def\wJ{\widetilde{J}}
\def\wQ{\widetilde{Q}}
\def\wS{\widetilde{S}}

\def\be{\begin{equation}}
\def\ee{\end{equation}}
\def\arr{\begin{array}{rll}}
\def\ea{\end{array}}
\def\bea{\begin{eqnarray}}
\def\eea{\end{eqnarray}}

\def\ic{{\rm i}}
\def\eu{{\rm e}}
\def\N2{$N{=}2$}
\def\diff{{\rm d}}

\def\sfrac#1#2{{\textstyle\frac{#1}{#2}}}
\def\>{\rangle}
\def\<{\langle}
\def\+{\dagger}
\def\={\ =\ }
\def\und{\qquad\textrm{and}\qquad}

\begin{document}
\renewcommand{\thefootnote}{\fnsymbol{footnote}}
\begin{titlepage}
\setcounter{page}{0}
\begin{flushright}
hep-th/0607215\\
ITP--UH--16/06\\
\end{flushright}

\vskip 3.0cm

\begin{center}

{\LARGE\bf Calogero models and }\\
\bigskip
{\LARGE\bf nonlocal conformal transformations}\\
\bigskip

\vskip 1cm

$
\textrm{\Large Anton Galajinsky\ }^{a} ,\quad
\textrm{\Large Olaf Lechtenfeld\ }^{b} ,\quad
\textrm{\Large Kirill Polovnikov\ }^{a}
$

\vskip 0.7cm
${}^{a}$ {\it 
Laboratory of Mathematical Physics, Tomsk Polytechnic University, \\
634050 Tomsk, Lenin Ave. 30, Russian Federation} \\
{Emails: galajin, kir @mph.phtd.tpu.edu.ru}

\vskip 0.4cm
${}^{b}$ {\it 
Institut f\"ur Theoretische Physik, Leibniz Universit\"at Hannover,\\
Appelstrasse 2, D-30167 Hannover, Germany} \\
{Email: lechtenf@itp.uni-hannover.de}

\vskip 0.2cm

\end{center}

\vskip 1cm

\begin{abstract} \noindent
We propose a universal method of relating the Calogero model to a set of 
decoupled particles on the real line, which can be uniformly applied to 
both the conformal and nonconformal versions as well as to supersymmetric 
extensions. For conformal models the simplification is achieved at the 
price of a nonlocal realization of the full conformal symmetry in the 
Hilbert space of the resulting free theory. As an application, we construct 
two different $N{=}2$ superconformal extensions.
\end{abstract}

\vspace{0.5cm}

PACS: 04.60.Ds; 11.30.Pb; 12.60.Jv\\ \indent
Keywords: Calogero model, nonlocal conformal transformations, supersymmetry

\end{titlepage}

\renewcommand{\thefootnote}{\arabic{footnote}}
\setcounter{footnote}0

\noindent
{\bf 1. Introduction}\\

\noindent
The range of physical and mathematical applications of the Calogero 
model is impressive. Being originally formulated as an exactly 
solvable multi-particle quantum mechanics in one 
dimension \cite{cal}, it played an important role in
the study of matrix models \cite{j,po}, 
fractional statistics \cite{poli}, classical and quantum 
integrable systems \cite{per}, 
the quantum Hall effect \cite{az}, 
superstring theory on the $AdS_2$ background \cite{ver}, 
the WDVV equation \cite{bgl} and BPS operators in $N{=}4$ SYM 
theory~\cite{pol2} (for a recent review see \cite{ppp}).

If one is concerned with only the pairwise interaction
$g^2{\sum}_{i<j}{(x^i-x^j)}^{-2}$ and disregards the harmonic potential
$\textstyle{ \om^2 {\sum}_{i<j}{(x^i-x^j)}^2}$, the Calogero model exhibits 
conformal symmetry \cite{bar}. This property and the fact 
that the isometry group of $AdS_2$ space is SO(1,2) led the authors of 
\cite{gib} to conjecture that an $N=4$ superconformal extension 
of the Calogero model might provide a microscopic description of the
extreme Reissner-Nordstr\"om black hole in the near horizon limit, 
which corresponds to $AdS_2 \times S^2$ geometry. 
Unfortunately, a consistent $N{=}4$ 
superconformal generalization of the Calogero model has not 
yet been constructed (for previous attempts see 
\cite{wyl,gal,bgk,bgl}). The latter problem 
partially motivated the present investigation.

It has been known since the original work of Calogero \cite{cal} that
in the presence of harmonic forces the energy eigenvalues of the problem 
differ from those of decoupled oscillators only by a constant. 
An explicit but non-unitary similarity transformation connecting 
their Hamiltonians has been constructed in \cite{gur}
(see also \cite{pgp} for a supersymmetric extension). 

When the harmonic potential is switched off one expects a similar  
relation between identical particles interacting via the inverse-square 
potential and free particles in one dimension to hold. A unitary 
transformation that maps the Hamiltonian of the Calogero model to that of 
free particles was constructed in \cite{pol}. However, the full 
conformal symmetry, which characterizes the case at hand, was not taken into
account. Note also that the transformation considered in \cite{pol} 
can not be obtained from that examined in \cite{gur} by taking 
the limit $\om \rightarrow 0$. This indicates that 
the two approaches are essentially different.

The purpose of this letter is to propose a universal method of relating the 
Calogero model to decoupled particles, which can be uniformly 
applied to both the conformal and nonconformal versions as well as to 
supersymmetric extensions. Our approach is different from
\cite{pol} in that it makes use of {\it all\/} conformal generators when 
constructing the transformation.
In other words, we study the behaviour of the Calogero model under 
specific (unitary) transformations generated by the conformal algebra so(1,2). 
As shown below, although the Hamiltonian~$H$
of the Calogero model can indeed be mapped to the free Hamiltonian $H_0$,
the generator~$K$ of special conformal transformations gets modified and 
keeps track of the original potential $H_{\rm int}=H{-}H_0$ 
via a nonlocal contribution,
\be
K \= \sfrac12 x^i x^i \qquad\longrightarrow\qquad
\wK \= K\ +\ \a^2 \eu^{\ic B} H_{\rm int}\,\eu^{-\ic B}
\qquad\textrm{with}\qquad
H_{\rm int}\= \sum_{i<j}\sfrac{g^2}{{(x^i-x^j)}^2}\ .
\ee 
Here $\a$ is a constant, and the explicit form of the operator $B$ 
is given below. A similar relation holds for an \N2 
superconformal extension of the Calogero model, for which also 
the superconformal generators are modified appropriately. 
Thus, after applying a unitary 
transformation one arrives at free particles in one dimension 
with the (super)conformal group being realized in a nonstandard 
(nonlocal) way. Although quantum states look particularly 
simple in this framework, the action of the full conformal group in the 
Hilbert space proves to be rather involved.

The organization of the paper is as follows. In sect.~2 we use 
general properties of the so(1,2) algebra and construct a novel unitary 
transformation which maps the conformal Calogero model to a set of 
free particles on the real line.
In sect.~3 the method is applied to the 
nonconformal Calogero model which features an external harmonic potential 
for each particle. A map to a set of decoupled harmonic oscillators is 
constructed and shown to be much simpler than the one proposed 
in \cite{gur}. We then proceed to explore supersymmetric generalizations
in sect.~4. The \N2 superconformal extension of the 
Calogero model built in \cite{fm} is related to a 
set of free \N2 superparticles, with the SU$(1,1|1)$ symmetry group 
being realized in a nonstandard fashion. We argue that the \N2
superconformal extension is not unique.
Furthermore, our transformation may pave the way to constructing $N{>}2$
superconformal extensions of the Calogero model from a set of free 
superparticles. We conclude by discussing possible further 
developments in sect.~5.  

\vspace{0.5cm}

\noindent
{\bf 2. From the Calogero model to free particles}\\

\noindent
Our starting point is the so(1,2) algebra realized in
the quantized $n$--particle Calogero model via the Weyl-ordered generators
\be\label{calogero}
H\=\sfrac{1}{2} p_i p_i+\sum_{i<j}\sfrac{g^2}{{(x^i-x^j)}^2} \ , \qquad 
D\=-\sfrac{1}{4}(x^i p_i +p_i x^i) \ , \qquad
K\=\sfrac{1}{2} x^i x^i \ ,
\ee
\be\label{algebra}
\textrm{which satisfy} \qquad
[H,D]=\ic H\ , \qquad [H,K]=2\ic D\ , \qquad [D,K]=\ic K\ . 
\ee
Here, $g$ is a dimensionless coupling constant ($[x]=[t^{\frac{1}{2}}]$), 
and the index $i$ labels $n$ identical particles (of unit mass) 
on the real line mutually interacting via the inverse-square potential.
Putting $g{=}0$ yields a free-particle representation of so(1,2), 
whose generators we denote by $H_0$, $D$ and $K$.

Each generic Lie-algebra element
\be\label{a}
A\=\a H + \b K +\g D \ ,
\ee
where the real constants $\a$ and $\b^{-1}$ have the dimension of length 
and $\g$ is dimensionless,
determines a unitary transformation 
\be
(H,D,K)\quad\longmapsto\quad(H',D',K') \= 
(\eu^{\ic A}H\eu^{-\ic A},\eu^{\ic A}D\eu^{-\ic A},\eu^{\ic A}K\eu^{-\ic A})
\ee
which is an automorphism of the algebra.
It is instructive to use the Baker--Campbell-Hausdorff formula
\be\label{bak}
T'\ \equiv\ \eu^{\ic A}T\eu^{-\ic A}\=\sum_{n=0}^\infty\frac{\ic^n}{n!}T'_n\ ,
\qquad\textrm{where}\qquad T'_0=T \quad\textrm{and}\quad 
T'_n =\underbrace{[A,[A, \dots [A,T] \dots]}_{n~\rm times} \ ,
\ee
and calculate the first three terms of the transformed Hamiltonian,
\be
H'_0 \= H \ ,\qquad
\ic H'_1 \= 2\b D + \g H \ , \qquad 
\sfrac{\ic^2}{2!} H'_2 \= (\sfrac{\g^2}{2}{-}\a\b)H + \b^2 K + \b\g D\ .
\ee
Apparently, the particular choice
\be\label{g}
\g \= \pm 2\,\sqrt{\a\b} \qquad\textrm{for}\quad \a\b>0
\ee
produces $\sfrac{\ic^2}{2!}H'_2=\b A$ and $H'_{n>2}=0$,
terminating the series in (\ref{bak}) at the third step. In what follows,
we always adopt this choice.

The condition~(\ref{g}) also terminates the series for the transformed
dilatation and special conformal generators, so together we have
\be
\label{hn}
\begin{aligned}
&
H' \= (1{+}\g{+}\a\b)H + \b(2{+}\g) D + \b^2 K \quad\
   \= \k_\pm^2 H + 2\b\k_\pm D + \b^2 K \ , \\[2pt]
&
D' \= -\a(1{+}\sfrac{\g}{2})H +(1{-}\sfrac{\g^2}{2})D+\b(1{-}\sfrac{\g}{2})K
   \= -\a\k_\pm H + (1{-}2\a\b)D + \b\k_\mp K \ , \\[2pt]
&
K' \= \a^2 H + \a(\g{-}2)D + {(1{-}\sfrac{\g}{2})}^2 K \qquad\quad
   \= \a^2 H - 2\a\k_\mp D + \k_\mp^2 K \ ,
\end{aligned}
\ee
where we abbreviated
\be
\k_\pm \ :=\ 1\pm\sqrt{\a\b} \qquad\textrm{for}\quad \a\b>0 \ .
\ee
An important simplification occurs for
\be
\label{r}
\a\b \= 1 \qquad\longrightarrow\qquad \k_+=2\ ,\quad \k_-=0
\ee
and the lower sign choice, $\g=-2$, namely
\be\label{tr}
H' \= \b^2 K \ ,\qquad
D' \= -D+2\b K\ , \qquad 
K' \= 4K-4\a D +\a^2 H \ .
\ee
Note that $H$ is mapped to the free-field generator $K\equiv K_0$.
For the upper sign choice one gets $K'=\a^2 H$ instead.

In our consideration it is only the structure of the conformal algebra 
which matters. So, by changing the operator $A$ in (\ref{a}) for
\be\label{b}
B \=\la H_0 +\s K +\d D \ ,
\ee
analogous relations hold for a system of free particles with the 
generators $H_0$, $D$ and $K$.
This observation suggests (in this respect see also \cite{pol})
that one can compose the transformations generated by $A$ and by $B$ to map
\be
H\ \mapsto\ K{\equiv}K_0\ \mapsto\ H_0 \qquad\textrm{via}\qquad
\a\b=1\ ,\quad \g=-2\ ,\quad \la\s=1\ ,\quad \d=+2\ .
\ee
The second map,
\be
(H_0,D,K)\quad\longmapsto\quad(H_0'',D'',K'') \=
(\eu^{\ic B}H_0\,\eu^{-\ic B},\eu^{\ic B}D\eu^{-\ic B},\eu^{\ic B}K\eu^{-\ic B})
\ ,
\ee
reads
\be\label{tr1}
K'' \= \la^2 H_0 \ , \qquad 
D'' \= -D-2\la H_0\ , \qquad
H_0'' \= 4 H_0 +4\s D +\s^2 K \ .
\ee
A successive application of the two transformations then produces
\be
H\quad\mapsto\quad\wH \= H_0\ , \qquad
D\quad\mapsto\quad\wD \= D \ , \qquad
K\quad\mapsto\quad\wK \= K + \a^2 H''_{\rm int}\ ,
\ee
provided we impose the further relations
\be\label{rrr}
\b\la\=-1 \qquad\Longrightarrow\qquad \a\s\=-1 \und \a+\la\=0\ .
\ee
Thus, with the help of the unitary operator $\eu^{\ic B} \eu^{\ic A}$ one can
transform the Hamiltonian of the Calogero model into that describing a system 
of free particles.

A few comments are in order. 
Firstly, a similar transformation of $H$ to $H_0$ has been discussed 
in \cite{pol}.
However, the authors of \cite{pol} employed (\ref{a}) with $\g=0$, whence
their Baker--Campbell-Hausdorff series did not terminate.
As was demonstrated above, our generic choices for $A$ and $B$ allow for
a drastic simplification.
Secondly, not the entire so(1,2) algebra was studied in \cite{pol}. 
According to our analysis,
the operator of special conformal transformations gets modified. In fact, it 
effectively ``hides'' the interaction potential,
which disappears for the Hamiltonian but gives a nonlocal contribution 
$\a^2 H''_{\rm int}=
\a^2\eu^{\ic B}\bigl(\sum_{i<j}\sfrac{g^2}{(x^i-x^j)^2}\bigr)\,\eu^{-\ic B}$
to $K$.
Thirdly, consistency requires the operator
$\eu^{\ic B} \eu^{\ic A}$ to be independent of the remaining free 
parameter $\a$, as the latter is not fixed by the formalism and has 
a dimension of length. In order to check this, let us differentiate 
$\eu^{\ic B} \eu^{\ic A}$ with respect to $\a$ and demonstrate that
\be\label{disare}
\sfrac{\diff}{\diff\a} \left( \eu^{\ic B} \eu^{\ic A}\right)\=0\ 
\ee
for our special Lie-algebra elements
\be \label{standardform}
A \= \a H + \sfrac{1}{\a}K - 2D \und
B \= -\a H_0 - \sfrac{1}{\a}K + 2D \ .
\ee
Taking into account also the commutation relations (\ref{algebra}), 
which are valid for both $H$ and $H_0$,
one can easily verify the relations
\be
\begin{aligned}
\bigl[\sfrac{\diff B}{\diff\a}, B^n\bigr] \= +2\ic n \sfrac{1}{\a} B^n 
&\qquad \Longrightarrow \qquad 
\sfrac{\diff\eu^{\ic B}}{\diff\a} \= 
\ic\bigl(\sfrac{\diff B}{\diff\a}+\sfrac{1}{\a}B\bigr) \eu^{\ic B} \=
2\ic\bigl(\sfrac{1}{\a}D-H_0\bigr) \eu^{\ic B}\ , \\[4pt]
\bigl[\sfrac{\diff A}{\diff\a}, A^n\bigr] \= -2\ic n \sfrac{1}{\a} A^n 
&\qquad \Longrightarrow \qquad
\sfrac{\diff\eu^{\ic A}}{\diff\a} \= 
\ic\bigl(\sfrac{\diff A}{\diff\a} -\sfrac{1}{\a}A\bigr) \eu^{\ic A} \=
2\ic\bigl(\sfrac{1}{\a}D-\sfrac{1}{\a^2}K\bigr) \eu^{\ic A}\ .
\end{aligned}
\ee
Together with (\ref{tr1}) they lead
to the desired result (\ref{disare}).

To summarize, the quantum mechanical Hamiltonian of the Calogero model can be 
transformed into a free Hamiltonian by applying an appropriate unitary 
transformation. Knowing its explicit form, the stationary states of the former
model can be immediately constructed from those of the latter. 
This is in agreement with the claim of \cite{poli} that the quantum Calogero 
model hiddenly describes free particles in one dimension. 
It should be remembered, however, that the price paid for this change of 
variables is a nonlocal realization of the full conformal algebra in the 
Hilbert space.

\vspace{0.5cm}

\noindent
{\bf 3. Adding the harmonic potential}\\

\noindent
Let us now add an external harmonic potential to the model.
The analysis of the previous section makes it clear that
our technique can still be applied.
Such a treatment of the Calogero model in the presence of a harmonic force
should be much less intricate than the computation of \cite{gur},
whose similarity transformation to decoupled harmonic oscillators
explicitly involves the correlated ground state of the Calogero model.

Consider then the Hamiltonian
\be
H_1\=\sfrac{1}{2} p_i p_i+\sum_{i<j}\sfrac{g^2}{{(x^i-x^j)}^2}+
\sfrac{\om^2}{2} x^i x^i \= H +\om^2 K \ .
\ee
Application of the first transformation with $A$ as in (\ref{hn}) for
the lower sign choice in~(\ref{g}) yields 
\be
H'_1 \= (\k_-^2 + \a^2\om^2) H\ +\ 
        (2\b\k_- - 2\a\k_+\om^2) D\ +\
        (\b^2 + \k_+^2\om^2) K \ .
\ee
It is clear that the first term on the r.h.s.\ can no longer vanish
for a real value of $\k_\pm=1\pm\sqrt{\a\b}$. Hence, we must allow
$\a$ and/or $\b$ to become complex in
\be\label{gg}
\k_- \= \ic\a\om \qquad\Longrightarrow\qquad
\a\b \= (1 - \ic\a\om)^2 \ ,
\ee
where $\a$ remains arbitrary. 
This means that, as in \cite{gur}, an ultimate 
similarity transformation is realized by a nonunitary operator.
With the above relations replacing (\ref{r}), the transformation
specializes to
\be
H' \= 2\ic\om D + (\sfrac{1}{\a^2}{-}4\ic\sfrac{\om}{\a}{-}2\om^2) K \ , 
\ee
which indeed reduces to (\ref{tr}) for $\om\to0$.

The same recipe works for the $B$ transformation, which is again found 
from $A$ by replacing $H\to H_0$ and changing the overall sign,
\be
\begin{aligned}
& A\=\phantom{-}\a H\,+\,\sfrac{1}{\a}(1{-}\ic\a\om)^2 K-2(1{-}\ic\a\om)D\ ,
\\[2pt]
& B\= -\a H_0 - \sfrac{1}{\a}(1{-}\ic\a\om)^2 K + 2(1{-}\ic\a\om)D\ .
\end{aligned}
\ee
It is straightforward to write down the second transformation and verify that
\be
\wH_1 \ \equiv\ \eu^{\ic B} \eu^{\ic A} H_1 \,\eu^{-\ic A} \eu^{-\ic B}
\= H_0 + \om^2 K \ ,
\ee
which proves that we have indeed mapped the nonconformal Calogero model 
to decoupled harmonic oscillators, via a simple explicit albeit non-unitary 
similarity transformation. Clearly, the limit $\om\to0$ connects with the
results of the previous section.

Finally, like in the previous case one can establish the independence
of the transformation on the parameter~$\a$.
Thus, the formalism developed in the preceding section is
universal and can be applied to both the conformal and 
nonconformal Calogero models.

\vspace{0.5cm}

\noindent
{\bf 4. Superconformal extensions}\\

\noindent
The unitary transformation constructed above has many interesting 
applications. In particular, it allows one to address the issue of 
superconformal extensions of the Calogero model. 
Below we treat in detail the \N2 case.
In our setting, this amounts to adding fermionic coordinates to the free model
and to properly modifying the nonlocal generator $\wK$ such as to close
the superconformal algebra. The inverse unitary transformation with the
standard form (\ref{standardform}) for $A$ and~$B$ then maps
the set of free superparticles back to the desired superconformal Calogero
model with the standard representation of~$K$.

Apart from the so(1,2) generators, the \N2 superconformal algebra contains
two supersymmetry generators $Q$ and $\bar Q$ which are 
hermitian conjugates of each other, two superconformal 
generators $S$ and $\bar S$ also related by hermitian conjugation, 
and a u(1) generator $J$.
Altogether there are four bosonic and four fermionic 
operators, which obey the nonvanishing commutation relations  
(suppressing hermitian conjugates)
\be \label{N2algebra}
\begin{aligned}
&
[H,D]=\ic H\ , \qquad\ \ 
[K,D]=-\ic K\, \qquad\
[Q,D]=\sfrac{\ic}{2} Q\ , \qquad\
[S,D]=-\sfrac{\ic}{2} S\ ,\\[2pt]
&
[Q,J]=-\sfrac{1}{2} Q\ , \qquad
[S,J]=-\sfrac{1}{2}S\ , \qquad
[H,K]=2\ic D\ , \qquad
[Q,K]=-\ic S\ , \\[2pt]
&
\{Q,\bar Q\}=2H\ , \qquad 
\{S,\bar S\}=2K\ , \qquad\,
\{Q,\bar S\}=-2D-2\ic J+\ic C\ , \qquad 
[S,H]=\ic Q\ .
\end{aligned}
\ee 
Here, $C$ is a real constant which stands for a central charge. 
For the realization of this algebra we need to add to the coordinates $x^i$
the same number $n$ of canonical pairs
of free fermions $\p^i$ and $\bar\p^i$,
subject to the standard anticommutation relations
\be
\{\p^i, \bar\p^j\}\=\d^{ij} \und
\{\p^i, \p^j\}\=0\=\{\bar \p^i, \bar\p^j\}
\qquad\textrm{with}\quad ({\p^i})^{*}=\bar\p^i\ .
\ee

The algebra~(\ref{N2algebra}) suggests that
$\wH=H_0=\sfrac12 p^i p^i$ is accompanied by
\be
\wQ\=Q_0\=\p^i p^i \und \widetilde{\bar Q}\=\bar Q_0\=\bar\p^i p^i
\ee
and the dilatation and u(1) generators
\be
\wD \= D\= -\sfrac{1}{4}(x^i p_i +p_i x^i) \und
\wJ \= J\= \sfrac{1}{4} \bigl(\p^i \bar\p^i -\bar\p^i \p^i \bigr) \ .
\ee
The remaining (conformal) generators $\wK$, $\wS$ and $\widetilde{\bar S}$
are nonlocal but acquire the standard form in the interacting model,
\be
K\= \sfrac12 x^i x^i \und 
S\= \p^i x^i\ ,\qquad \bar S\=\bar\p^i x^i\ .
\ee

The goal is to construct the interacting-model Hamiltonian $H$ and supercharges
$Q$ and~$\bar Q$ by working our way back from the free model with the help of
the algebra~(\ref{N2algebra}).
To this end, we begin with the special conformal generator and
parametrize as before
\be \label{Kansatz}
\wK \= K\ +\ \a^2\,\eu^{\ic B} H_{\rm int}\,\eu^{-\ic B}
\qquad\textrm{but with}\qquad
H_{\rm int} \= {\sum}_{i<j}~\sfrac{g^2}{(x^i-x^j)^2} \ +\ V\ ,
\ee
allowing for a new contribution~$V$ due to the fermions.
The algebra commutators (\ref{N2algebra}) then consistently fix
the form of the superconformal generator $\wS$, 
\be \label{Stilde}
[\wQ,\wK]\=-\ic \wS \qquad\Longrightarrow\qquad 
\wS\=S - \ic\a\,\eu^{\ic B}\,[S,V]\,\eu^{-\ic B}\ .
\ee
Hermitian conjugation produces $\widetilde{\bar S}$. Other 
structure relations of the superconformal algebra~(\ref{N2algebra})
yield the following restrictions on the form of $V$:
\bea\label{str}
&&
[K,V]\=0\ , \qquad [D,V]\=-\ic V\ , \qquad [J,V]\=0\ ,
\nonumber\\[2pt]
&&
[Q,\,H_{\rm int}]\ +\ \ic[H_0{+}V,[S,V]]\=0\ ,\qquad
\{S,[\bar S,V]\}\=C\ ,
\\[2pt]
&&
\{[S,V],[\bar S,V]\}\ +\ \ic\{Q,[\bar S,V]\}\ +\ 
\ic\{\bar Q,[S,V]\}\ +\ 2 H_{\rm int}\=0\ ,
\nonumber
\eea 
plus their hermitian conjugates. 

Let us define an \N2 Calogero model
by finding a solution to the equations (\ref{str}). 
The first line in (\ref{str}) implies that the potential $V$ is a homogeneous 
function of the $x^i$ of degree $-2$. Being u(1) neutral, it involves an
equal number of $\p^i$ and $\bar\p^i$. Thus, it is natural to take the 
simplest ansatz 
\be \label{vansatz}
V\=V_{ij}(x)\,\p^i\bar\p^j \= 
\sfrac12 V_{ii}(x)\ +\ \sfrac12 V_{ij}(x)\,[\p^i,\bar\p^j]
\ee
with unknown functions $V_{ij}(x)$.
Substituting this form into the remaining (anti)commutators
in  (\ref{str}) one obtains a system of 
partial differential equations,
\be\label{vv}
\begin{aligned}
&
-2 V_{ij}\=\partial_i (V_{jp} x^p)\ +\ \partial_j (V_{ip} x^p)\ , \qquad 
V_{ij}\ +\ \partial_j(V_{ip} x^p)\=0\ , \qquad
\partial_p V_{ij}\=\partial_i V_{pj}, \\[2pt]
& \qquad\quad
\partial_i (V_{ip} x^p)\ +\ (V_{ip} x^p)(V_{is} x^s)\ -\ 2\, 
\textstyle{{\sum}_{i<j}~\frac{g^2}{{(x^i-x^j)}^2}}\=0\ , \qquad 
V_{ij} x^i x^j\=C\ .
\end{aligned}
\ee
The first equation implies that $V_{ij}=V_{ji}$. Then
the second restriction gives the condition
\be
\partial_i (V_{jp} x^p)\ -\ \partial_j (V_{ip} x^p) \=0 
\qquad\Longrightarrow\qquad
V_{ip} x^p\= \partial_i \Phi
\ee
with some scalar function $\Phi$. The remaining equations in (\ref{vv})
imply that
\be \label{vij}
V_{ij}\=-\partial_i \partial_j \Phi
\ee
and constrain $\Phi$ to obey the partial differential equations
\be\label{pe}
\partial_i \partial_i \Phi\ +\ (\partial_i \Phi)(\partial_i \Phi) \=
2\,\textstyle{{\sum}_{i<j}~\frac{g^2}{{(x^i-x^j)}^2}} \und 
x^i \partial_i \Phi \= C \ .
\ee
Any solution $\Phi$ to these equations will give rise to an \N2 superconformal
extension of the Calogero model. 

The general solution to (\ref{pe}) can be put in the form
\be\label{phi}
\Phi\=\mu\,{\sum}_{i<j}\ln|x^i-x^j|\ +\ \nu\,\ln\sqrt{x^2}
\ +\ \Lambda\bigl(\{\sfrac{x^i}{x^1}\}\bigr)\ ,
\ee
where $\mu$ and $\nu$ are dimensionless constants, $x^2\equiv x^i x^i$,
and $\Lambda$ is a general function of coordinate ratios. Putting for
simplicity $\Lambda\equiv0$ and inserting (\ref{phi}) into~(\ref{pe}), we find
the conditions 
\be\label{supp}
\m\,(\m{-}1)\=g^2\ >-\sfrac14 \und \n\,\bigl(\n+n(n{-}1)\,\m+n-2\bigr)\=0 \ ,
\ee
which give four solutions for the pair $\mu(n,g)$ and $\nu(n,g)$.
The central charge is fixed at
\be
C(n,g) \= \sfrac{n(n-1)}{2}\,\mu \ +\ \nu\ .
\ee
Differentiating twice as in~(\ref{vij}) and inserting in~(\ref{vansatz}) yields
\be
V \= \sum_{i<j}~\sfrac{\mu}{(x^i-x^j)^2}\ -\ \sfrac{n-2}{2}\,\sfrac{\n}{x^2}\
+\ \sfrac12\sum_{i\neq j}~\sfrac{\mu}{(x^i-x^j)^2}\ [\p^i,\bar\p^i{-}\bar\p^j]\
-\ \sfrac12\sum_{i,j}~\sfrac{\n}{x^2}\,\sfrac{x^2\d^{ij}-2x^ix^j}{x^2}\ 
[\p^i,\bar\p^j]
\ee
and, hence, with (\ref{supp}) the interaction Hamiltonian
\be\label{n2h}
H_{\rm int} \= \sum_{i<j}~\sfrac{\m^2}{ {(x^i-x^j)}^2}\ +\ 
\sfrac12 \sum_{i\neq j}~\sfrac{\mu}{(x^i-x^j)^2}\ [\p^i,\bar\p^i{-}\bar\p^j]\ 
-\ \sfrac{n-2}{2}\,\sfrac{\n}{x^2}\ -\ \sfrac12\sum_{i,j}~
\sfrac{\n}{x^2}\,\sfrac{x^2\d^{ij}-2x^ix^j}{x^2}\ [\p^i,\bar\p^j]
\ee
but also $\wK$ and $\wS$.
The original Calogero coupling~$g^2$ has been replaced by $\mu^2$,
of which $\nu$ is a function via (\ref{supp}).
By the very construction,
this $H=H_0+H_{\rm int}$ along with $D$ and $K$ from (\ref{Kansatz}) furnish a representation 
of so(1,2). Therefore, they can be used to construct the inverse 
transformation $\eu^{-\ic A} \eu^{-\ic B}$ and hence the supercharge,
which for $\nu{=}0$ reads
\be
Q\= \eu^{-\ic A}\eu^{-\ic B}\bigl(\p^i p^i\bigr)\,\eu^{\ic B}\eu^{\ic A}
\= \p^i p^i\ +\ \ic[V,S] \=
\p^i\bigl(p^i+\ic\textstyle{{\sum}_{k(\ne i)}~\frac{\m}{x^i-x^k}}\bigr)\ .
\ee
It may be checked that the same transformation maps $\wS$ of (\ref{Stilde})
back to~$S$ as it should.

Beautifully enough, with $\n{=}0$ we have reproduced precisely the 
\N2 superextension constructed by Freedman and Mende~\cite{fm} in 
the framework of supersymmetric quantum mechanics.
For the other solution to (\ref{supp}), $\n=2-n-n(n{-}1)\m$, 
we have apparently found an alternative superextension 
(see also~\cite{gh}).

\vspace{0.5cm}

\noindent
{\bf 5. Concluding remarks}\\

\noindent
In this paper we have constructed a simple unitary transformation
relating the conformal Calogero model to a system of free particles on 
the real line. The simplification was achieved at a price of a highly 
nontrivial and, in particular, nonlocal realization of the full 
conformal symmetry in the resulting free theory. 
The transformation was shown to be universal and 
applicable to the nonconformal Calogero model as well as to \N2 
supersymmetric extensions. 
In the latter case we reconstructed not only the model of Freedman and Mende
but found a second variant.

Turning to possible further developments, first to mind comes the
$N{=}4$ superconformal extension of the Calogero model, which 
seems crucial for testing a conjecture of Gibbons and Townsend \cite{gib}. 
The construction realized for the su$(1,1|1)$ superalgebra in sect.~4 can 
literally be generalized to the su$(1,1|2)$ superalgebra. 
This project is under way.
Another interesting point is to employ our transformation for deriving
the propagator of the Calogero model starting from the free propagator.
Finally, it may be worthwhile to generalize the analysis of sect.~3 
to the case of a harmonic pair potential.

\vspace{0.5cm}

\noindent{\bf Acknowledgements}\\

\noindent
A.G.\ thanks the Institute for Theoretical Physics at Hannover University 
for the hospitality extended to him during the course of this work.
He is grateful to the DAAD for financial support (A/06/27302). 
The research was supported by RF Presidential grants MD-8970.2006.2, 
NS-4489.2006.2, INTAS grant 03-51-6346, DFG grant 436 RUS 113/669/0-3, 
RFBR grant 06-02-16346 and RFBR-DFG grant 06-02-04012.

\end{document}